\documentclass[14pt]{article}
\usepackage{amssymb}
\usepackage{amsmath,bm}
\usepackage{array}

\title{Could the real (not virtual) static observer exist outside a Schwarzschild black hole?}
\author{Victor Berezin}
\date{
\small Institute for Nuclear Research, Russian Academy of Sciences,
\\ 60th October Anniversary pr., 7-a, 117312 Moscow, Russia \\
{ \ } \\
{ \ } \\
e-mail: berezin@ms2.inr.ac.ru
}

\begin{document}

\maketitle

\vglue 2cm

\begin{abstract}
The aim of this Letter is rather pedagogical. We considered the static spherically symmetric ensemble of observers, having finite bare mass and trying to measure geometrical and physical properties of the environmental static (Schwarzschild) space-time. It is shown that, using the photon rockets (which the mass together with the mass of their fuel is also taken into account) they can managed to keep themselves on the fixed value of radius. The process of diminishing the total bare mass up to zero lasts infinitely long time. It is important that the problem is solved self-consistently, i.e., with full  account for the back reaction of both bare mass and radiation from rockets on the space-time geometry.
\end{abstract}

\newpage

\section{Introduction}

Everybody knows what the important role is played by test particles in General Relativity. It is supposed they have infinitesimal masses and sizes, because only for such small and almost weightless bodies the Einstein's equivalence principle is valid. The finite masses and sizes would inevitably give rise to he self-interaction and gravitational radiation, hence disturbing the geometry under the test. This is clearly possible for freely falling observers that are moving along time-like geodesics providing us with the very important information about the curved backgrounds.

But the situation is not so evident in the case of static observer, for example, outside a Schwarzschild black hole. The problem is that the static observer in the static gravitational field undergoes a uniform acceleration. To produce such an acceleration, one needs a force acting for (at least) some period of time. To produce such a force, one needs some energy. Such an energy weighs and produces its own gravitational field. Moreover it requires some container. To keep static such a container filled with a fuel, one needs a force and so on, and so forth. The most efficient device for converting a fuel into the reactive force is a photon rocket. In a sense, our model represents a spherically symmetric realization of an ensemble of Kinnersly's photon rockets \cite {K}. We need spherical symmetry in order to be sure that the geometry outside the layer with our observers equipped with photon rockets is that outside of a Schwarzschild black hole. To make the problem exactly solvable we assume that such a layer is infinitely thin.

\section{Constructing the model}

Our model consists of three pieces. The first one is a part of the Schwarzschild space-time with given mass $m_{out}$, that should be tested by an ensemble of static observers. The second is a thin shell where the observers equipped with photon rockets are situated. We demand that the shell is kept at the constant value of radius, $r_0$, by the reaction force produced by radiation coming from these rockets. The third piece is the space-time inside the thin shell. Since the rays of light are going towards the center of the shell sphere, we have there an advanced Vaidya metric.

The famous Schwarzschild metric \cite{Sch} outside the corresponding event horizon (Schwarzschild radius $r > r_g = 2 \, G m$)
\begin{eqnarray}
\label{schmetr}
ds^2 = F dt^2 - \frac{1}{F}dr^2 - r^2 (d\vartheta^2 + \sin^2{\vartheta} d\varphi^2) \, , \nonumber \\
F = 1 - \frac{2 \, G\, m}{r}, \;\;\;\; m = const
\end{eqnarray}
( $r$ is the radius of the sphere, $\vartheta$ and $varphi$ are usual spherical angles, and we put the speed of light $c = 1$) can be put in another form by using retarded, ($u,r$), or advances, ($v,r$), Finkelstein coordinates. It is convenient to introduce the common notations: $z = t - \varepsilon r^{\star}$, where $z = u$ for $\varepsilon = +1$ and $z = v$ for $\varepsilon = -1$, and $r^{\star}$ is called the "tortoise" coordinate, it is defined by the relation $dr^{\star} = \sigma \frac{dr}{F}$, with $\sigma$ is yet another sign function. The latter shows whether radii increase in the outer normal direction to the shell with observers $(\sigma = +1)$, or they decrease $(\sigma = -1)$. In the Finkelstein coordinates $(z,r)$ the Schwarzschild metric takes the form
\begin{equation}
\label{Fink}
ds^2 = F dz^2 + 2 \varepsilon \sigma dz dr -r^2 (d\vartheta^2 + \sin^2{\vartheta} d\varphi^2).
\end{equation}
The Vaidya metric \cite{V} differs from the Schwarzschild one by allowing the mass $m$ in the Finkelstein form of the metric to be a function of the coordinate $z$ (i.e., of the retarded, $u$, or advanced, $v$, time). So, $F = 1 - \frac{2 \, G\,m(z)}{r}$. It is easy to check that the energy-momentum tensor $T_{\mu, \nu}$ in the Vaidya space-time equals
\begin{eqnarray}
\label{Vemt}
T_{zz} &=& - \frac{\varepsilon \sigma}{4\, \pi r^2} \frac{dm}{dz}\, , \nonumber \\
T_{zr} = T_{rr} &=& T_{22} = T_{33} = 0 \, .
\end{eqnarray}

To obtain the complete geometry we have to match the interior and exterior metriñs on the thin shell (with observers), thus relating their parameters. The matching conditions are known as Israel equations \cite{I}. They connect the surface energy-momentum tensor $S^j_i$ of the shell to the jump in the extrinsic curvature tensor $K^j_i$ describing embedding of his shell into exterior $(out)$ and interior $(in)$ manifolds. To write down these equation, let us introduce the Gaussian normal coordinates associated with the shell (which is time-like in our case):
\begin{equation}
\label{Gnc}
ds^2 = -dn^2 + \gamma_{ij}(n,x) dx^i dx^j
\end{equation}
The shell is situated at $n = 0$, where $n$ is the spatial coordinate normal to the shell ($n < 0$ inside and $n> 0$ outside). The above-mentioned extrinsic curvature tensor is defined then as $K_{ij} = - \frac{1}{2} \gamma_{ij,n} \; (i,j = 0,2,3)$ - comma denotes a partial derivative. The surface energy-momentum tensor of the shell is defined by $T^{\nu}_{\mu} = S^{\nu}_{\mu} \delta (n) + ... (\mu, \nu = 0,1,2,3)$, $\delta ()$ - Dirac function. The metric coefficients $\gamma_{\mu \nu}(n,x)$ are assumed continuous on the shell, so, if some of their the first derivatives undergo jumps at the shell position, the corresponding second derivatives contained in the Einstein equations, will exhibit the $\delta$-function behavior. Integration across the shell gives us $S^n_n = 0$ and $S^n_i = 0$, that can be viewed as the shell definition, and the following Israel equations:
\begin{equation}
\label{Iseq}
- \left[K^j_i \right] + \delta^j_i \left[K \right] = 8 \, \pi \, G \,S^j_i \, ,
\end{equation}
where $G$ is the Newtonian constant, $K = K^l_l\, , \; \delta^j_i$ is the unit tensor, and $[\cdots ] = (out) - (in)$ stands for the jump in the corresponding quantity. In many cases it is appeared convenient to add the continuity equation for $S^j_i$:
\begin{equation}
\label{conteq}
S^j_{i|j} + \left [T^n_i \right] = 0 \, ,
\end{equation}
that follows from the Bianci identities (the vertical line denotes covariant differentiation with to the metric on the shell, $\gamma _{ij}$) and is, infact, the differential consequence of the Israel equations and Einstein equations on both sides of the shell.

In the case of spherical symmetry everything is very much simplified. Namely, we have
\begin{eqnarray}
\label{ss}
ds^2 &=& g_{00}(t,n)\, dt^2 - dn^2 - R^2(t,n) (d\vartheta^2 + \sin^2{\vartheta} d\varphi^2) \, , \nonumber \\
K^j_i &=& \left (K^0_0\, , \; K^2_2 = K^3_3 \right) \, , \nonumber \\
S^j_i &=& \left(S^0_0\, , \; S^2_2 = S^3_3 \right) \, ,
 \end{eqnarray}
 and the Israel equations are reduced to
 \begin{eqnarray}
 \label{sphsie}
 \left [K^2_2 \right] &=& 4 \,\pi \, G \, S^0_0 \, , \nonumber \\
 \left [K^0_0 \right] + \left [K^2_2 \right] &=& 8 \, \pi \, G \, S^2_2 \, .
 \end{eqnarray}
 To proceed, let us introduce the proper time $\tau$ for the observers sitting on the shell, by $d\tau = g_{00}(t,0) dt$, and the invariant $\Delta = \gamma^{ik} R_{,i} R_{,k}$ which is, together with the invariant radius $R$, defines the global geometry of a generic spherically symmetric space-time \cite{BKT}, and let $\rho(\tau) = R(t,0)$. Clearly, in Gausian normal coordinates, $\Delta = \frac{1}{g_{00}}R_{,t}^2 - R_{,n}^2$. This allows us to calculate easily $K^2_2$. Indeed, $K_{22} = R R_{,n}\, , \; K^2_2 = - \frac{R_{,n}}{R}$, so, on the shell (dot means the proper time derivative)
 \begin{equation}
 \label{k22}
 K^2_2 = - \frac{R_{,n}}{R} = - \frac{\sigma}{\rho} \sqrt{\dot \rho^2 - \Delta}\,.
 \end{equation}
 The sign function $\sigma$ is the same as was introduced when defining a "tortoise' coordinate $r^{\star}$, therefore $\frac{\partial r^{\star}}{\partial n} > 0$. The calculation of $S^0_0$ is much more cumbersome, we will not show it here. By using Einstein equations on both sides of the shell, the Israel equations can be put into the physically meaningful form
 \begin{eqnarray}
 \label{finisreq}
 \sigma_{in} \sqrt{\dot\rho^2 - \Delta_{in}} - \sigma_{out} \sqrt{\dot\rho^2 - \Delta_{out}} = 4 \, \pi \, \rho \, S^0_0 \, ,\nonumber \\
 \frac{\sigma_{in}}{\sqrt{\dot\rho^2 - \Delta_{in}}} \left(\ddot\rho + \frac{1 + \Delta_{in}}{2 \, \rho } - 4 \, \pi \, G \, \rho \, T^n_n(in) \right)\nonumber \\
  - \frac{\sigma_{out}}{\sqrt{\dot \rho^2 - \Delta_{out}}} \left(\ddot \rho + \frac{1 + \Delta_{out}}{2 \, \rho } - 4 \, \pi \, G \, \rho \, T^n_n(out) \right) = 4 \, \pi \, G \, \left(2 \,S^2_2 - S^0_0 \right) \, , \nonumber \\
 \dot S^0_0 + 2 \frac{\dot \rho}{\rho} \left(S^0_0 - S^2_2 \right) + \left[T^n_0 \right] = 0\, .
 \end{eqnarray}
 where we added the continuity equation for further convenience.

 Now, let us have a look at what else must be done.

 Fist, we should demand our observers to be fixed at some value of radius. Therefore, we put $\dot\rho = \ddot\rho = 0\, , \; \rho = r-0 = const$ in the Israel equations.

 Second, we have to specify he shell's equation of state. Since the observers, by definition, are independent of each other, the best approximation will be to consider a dust shell, for which $S^2_2 = 0$. Then, introducing the bare mass $M$ of the shell by $M = 4\, \pi\, r_0^2 \, S^0_0$ we obtain from the continuity equation
 \begin{equation}
 \label{barem}
 \dot M + 4 \, \pi \, r_0^2\, \left[T^n_0 \right] = 0 \, .
 \end{equation}

 Third, we know the form of the energy-momentum tensor inside the shell in the Finkelstein coordinates $(z,r)$ (outside it is simply zero). The same tensor enters Israel equations in the form $T_n^n$ and $T^n_0$ in the Gaussian normal coordinates. Thus, we should find the corresponding coordinate transformation. This is a rather simple exercise, below we present the result:
 \begin{eqnarray}
 \label{transf}
 \frac{\partial z}{\partial \tau} &=& \frac{\sqrt{\dot\rho^2 + F} - \varepsilon \sigma \dot\rho}{F} \, , \nonumber \\ \frac{\partial z}{\partial n} &=& \sigma \frac{\dot \rho - \varepsilon \sigma \sqrt{\dot\rho^2 + F}}{F} = - \varepsilon \frac{\partial z}{\partial \tau}
 \end{eqnarray}
 Thus, for the energy-momentum tensor we have
 \begin{eqnarray}
 \label{tnn}
 T^n_n = - \left(\frac {\sqrt{\dot\rho^2 + F} - \varepsilon \sigma \dot\rho}{F} \right)^2 T_{zz} \, , \nonumber \\
 T^n_0 = - \varepsilon T^n_n \, ,
 \end{eqnarray}
 or, after putting $\rho = r_0 = const$ and substituting for $T_{zz}$ the Eqn.(\ref{Vemt}),
 \begin{eqnarray}
 \label{fintnn}
 T^n_n &=& \frac{\varepsilon \sigma}{4 \, \pi \, r_0^2 \, F} \frac{dm}{dz} \, , \nonumber \\
 T^n_0 &=& - \frac{\sigma}{4 \, \pi \,r_0^2 \, F} \frac{dm}{dz} \, .
 \end{eqnarray}
 At last, we have everything at hand to answer the question in the title.

 First of all, we should say some words about the global geometry of the Schwarzschild space-time. At the moment of time symmetry (which can be considered as arbitrary), its spatial geometry represents the so called Einstein-Rosen bridge, two isometrical parts of which are causally disconnected, i.e., it is a geometry of non-traversable wormhole. On both sides of the bridge we have the asymptotically flat regions and the throat in-between. The size of the throat is just the gravitational radius of the Schwarzschild black hole. If we introduce some monotonic continuous radial coordinate,say, $n$in our case, that runs from $(-\infty)$ at the asymptotically flat region on one end of the Einstein-Rosen bridge to $(+\infty)$ - on the other end, then the radii first decrease from $(+\infty)$ to the value of gravitational radius $r_g$ at the throat, so $\sigma = -1$ there, then they start to increase on "our" side of the bridge, so $\sigma = +1$ here. We want that our observer would see the black hole inside, so we put $\sigma_{in} = +1$. And we consider two quite different physical situations - for $\sigma_{out} = +1$ and for $\sigma_{out} = -1$.

 Let us start with the case $\sigma_{out} = +1$ which is intuitively much more transparent. Our observers fell the gravitational attraction from inside only, so each of them should have only one photon rocket with its light rays streaming inside. This means that we should use the advanced Finkelstein coordinates inside and put there $\varepsilon = -1,; z = v$, so, $F_{in} = 1 - \frac{2\,G\,m(v)}{r_0}$. Outside we have the ordinary Schwarzschild solution with spatial infinity and fixed mass $m_{out} = m_0 = const$ and $F_{out} = F_0 = 1 - \frac{2\,G\,m_0}{r_0} = const$. Thus, $T^n_n (in) = T^n_0 (in) = - \frac{1}{4\, \pi\, r_0^2 \, F_{in}} \frac{dm}{dv} = \frac{\dot M}{4\, \pi\,r_0^2}, \;\; T^n_n (out) = T^n_0 (out) = 0$, and the Israel equations become
 \begin{eqnarray}
 \label{finsheq}
 \sqrt{F_{in}} - \sqrt{F_0} &=& \frac{G\,M(\tau)}{r_0} \, , \nonumber \\
 \frac{1}{F_{in}} \left( \frac{1 - F_{in}}{2\,r_0} - \frac{G}{r_0} \dot M \right) - \frac{1 - F_0}{2 \sqrt{F_0} r_0} &=& - \frac{G\,M}{r_0^2} \, .
 \end{eqnarray}
 Substituting $\sqrt{F_{in}}$ from the first of these equation into the second one, we get the simple ordinary differential equation for $M(\tau)$:
 \begin{equation}
 \label{eqM}
 2 \, r_0 \, \dot M = M \left( \frac{G\,M}{r_0} - A \right)\, ,
 \end{equation}
 where $A = \frac{1 - F_0}{\sqrt{F_0}}$. Its solution is
 \begin{equation}
 \label{solM}
 \frac{G\,M}{r_0} = \frac{A}{e^{\frac{A(\tau - \tau_0)}{2\,r_0}} + 1} \, ,
 \end{equation}
 $\tau_0$ being an integration constant. Introducing the initial value of bare mass $M_0 = M(\tau = 0), \; \frac{G\,M_0}{r_0} < \sqrt{F_{in}(\tau = 0)}$, we can rewrite the above expression as follows
 \begin{equation}
 \label{M0}
 M = \frac{M_0}{e^{\frac{A \tau}{2 r_0}} \left(1 - \frac{G\,M_0}{A\,r_0} \right) + \frac{G\,M_0}{A\,r_0}} \, .
 \end{equation}
 Of course, it is also possible to find $m_{in}(\tau)$ and $\tau (v)$ on the shell, but we will not do it here. The problem is already solved because we know now that it takes an infinite period of time to convert the whole bare mass of the static shell into the in-falling radiation. It is in this sense that the existence of a physical static observer outside a Schwarzschild black hole becomes possible.

 In the end, for the sake of completeness, let us discuss briefly the case $\sigma_{out} = -1$. Now our observers see two black holes, both inside and outside. And two throats belonging to two Einstein-Rosen bridges separate the observers from two asymptotically flat regions. There exists no physically acceptable solution to our equations for positive bare mass and outgoing radiation. This fact can be understood as follows. First, now "the inside" and "the outside" are on equal terms, the forces are attractive from both sides, so, to stay at the fixed value of radius, the observer should radiate photons in both directions. Let the black hole mass inside is less that that of outside. Then, to compensate the more powerful attraction, the observers have to enhance the radiation going outside, thus increasing more and more the larger black hole mass. The situation is obviously unstable. It is noteworthy to mention that in this case the black hole geometries out of the shell are no more static.
\section{Acknowledgments}

The author is greatly indebted to Alexei Smirnov and Vyacheslav Dokuchaev for valuable discussions.

This work was supported by the grant No.~10-02-00635-a from the Russian Foundation of Fundamental Investigations (RFFI).


\begin{thebibliography}{9999}
\bibitem{K}
W.Kinnersly. {\it Phys.Rev.} {\bf 186} (1969) 1335;
J.Podolsky. {\it IJMPD} {\bf 20} (2011) 335.

\bibitem{Sch}
K.Schwarzschild. {\it Sitzber Deut.Akad. Wiss. Berlin, Kl.Math.-Phys.
Thech.} (1916) 189.

\bibitem{V}
P.C.Vaidya. {\it Proc.Indian Acad.Sci.} {\bf A 33} (1951) 264.

\bibitem{I}
W.Israel. {\it Nuovo Cim.} {\bf B44} (1966) 1, {\bf B48} (1967) 463.

\bibitem{BKT}
V.A.Berezin, V.A.Kuzmin, I.I.Tkachev. {\it Phys.Rev.} {\bf D36} (1987) 2919.

\end{thebibliography}
\end{document}